\documentclass[fleqn,usenatbib]{mnras}

\usepackage{newtxtext,newtxmath}

\usepackage[T1]{fontenc}

\DeclareRobustCommand{\VAN}[3]{#2}
\let\VANthebibliography\thebibliography
\def\thebibliography{\DeclareRobustCommand{\VAN}[3]{##3}\VANthebibliography}

\usepackage{graphicx}	
\usepackage{amsmath}	
\usepackage{float}
\usepackage{subfigure}
\usepackage[section]{placeins}

\newcommand{\msol}{$\rm{M}_\odot$}

\newcommand{\jzjc}{$j_z/j_{\mathrm{circ}}$}
\newcommand{\jpjc}{$j_{\rm{p}}/j_{\mathrm{circ}}$}
\newcommand{\emin}{$e/e_{\mathrm{min}}$}
\newcommand{\kappaco}{$\kappa_{\mathrm{co}}$}

\newcommand{\meanjzjc}{$\langle$\jzjc$\rangle \,$}
\newcommand{\meanemin}{$\langle$\emin$\rangle \,$}

\newcommand{\eagle}{{\sc{Eagle}}}
\newcommand{\subfind}{{\sc{Subfind}}}

\newcommand{\fdisk}{$f_{\rm{disc}}$}

\newcommand{\fihl}{$f_{\rm{IHL}}$}

\newcommand{\rtrans}{$r_{\rm{IHL}}$}

\title[Identifying intra-halo light]{Identifying the discs, bulges, and intra-halo light of simulated galaxies through structural decomposition}

\author[Proctor et al.]{
Katy L. Proctor,$^{1,2}$\thanks{E-mail: katy.proctor@icrar.org}
Claudia del P. Lagos,$^{1,2}$
Aaron D. Ludlow$^{1}$
Aaron S. G. Robotham$^{1,2}$
\\
$^{1}$International Centre for Radio Astronomy Research (ICRAR), M468, University of Western Australia, 35 Stirling Highway, Crawley, WA 6009, Australia. \\
$^{2}$ARC Centre of Excellence for All Sky Astrophysics in 3 Dimensions (ASTRO 3D).
}

\date{Accepted XXX. Received YYY; in original form ZZZ}

\pubyear{2023}

\begin{document}
\label{firstpage}
\pagerange{\pageref{firstpage}--\pageref{lastpage}}
\maketitle

\begin{abstract}
We perform a structural decomposition of galaxies identified in three cosmological hydrodynamical simulations by applying Gaussian Mixture Models (GMMs) to the kinematics of their stellar particles. We study the resulting disc, bulge, and intra-halo light (IHL) components of galaxies whose host dark matter haloes have virial masses in the range $M_{200}=10^{11}$-- $10^{15}\,{\rm M_\odot}$. Our decomposition technique isolates galactic discs whose mass fractions, $f_{\rm disc}$, correlate strongly with common alternative morphology indicators; for example, $f_{\rm disc}$ is approximately equal to $\kappa_{{\rm co}}$, the fraction of stellar kinetic energy in co-rotation. The primary aim of our study, however, is to characterise the IHL of galaxies in a consistent manner and over a broad mass range, and to analyse its properties from the scale of galactic stellar haloes up to the intra-cluster light. Our results imply that the IHL fraction, \fihl, has appreciable scatter and is strongly correlated with galaxy morphology: at fixed stellar mass, the IHL of disc galaxies is typically older and less massive than that of spheroids. Above $M_{200}\approx 10^{13}\,{\rm M_\odot}$, we find, on average, $f_{\rm IHL}\approx 0.37$, albeit with considerable scatter. The transition radius beyond which the IHL dominates the stellar mass of a galaxy is roughly $30\,{\rm kpc}$ for disc galaxies, but depends strongly on halo mass for spheroids. However, we find that no alternative IHL definitions -- whether based on the ex-situ stellar mass, or the stellar mass outside a spherical aperture -- reproduce our dynamically-defined IHL masses.
\end{abstract}


\section{Introduction}\label{sec:intro}

In the `$\Lambda$-cold dark matter' ($\Lambda$CDM) cosmological model, structure forms in a hierarchical manner. 
Galaxies are expected to have undergone numerous accretion events over their lifetime, building up their stellar
mass at least in part through mergers with lower-mass ones. In principle, these accretion events should leave an
observable imprint on present-day galaxies in the form of an extended stellar component: the intra-halo light
\citep[IHL; see e.g.][]{purcell_shredded_2007}. The ubiquity of merger events in the $\Lambda$CDM model implies
that the IHL should be present, to some extent, in most galaxies \citep{bullock_tracing_2005}. Indeed, the IHL has
been observed across a broad range of galaxy masses; it is typically referred to as a stellar halo at the galactic
scale \citep[see][for a review of the Milky Way's, MWs, stellar halo]{helmi_stellar_2008}, and as intra-group (cluster)
light at the galaxy group (cluster) scale \citep[see][for reviews]{contini_origin_2021, montes_new_2022}.

Our understanding of the mass fraction, spatial distribution and stellar populations of the IHL, as well as how they
vary with host halo mass, is limited, partly due to difficulties defining the IHL unambiguously
\citep[see e.g.][and references therein]{sanderson_reconciling_2018}. Nonetheless, there is mounting evidence hinting
at a connection between present day IHL properties and the formation of its host galaxy and dark matter (DM) halo.
In the case of the MW, observations of individual halo stars \citep[e.g., by the][]{gaia_collaboration_gaia_2021}
have revolutionised our understanding of the Galaxy's formation history through the discovery of distinct
substructures in 6D phase space \citep[e.g., the Gaia Enceladus/Sausage;][]{belokurov_co-formation_2018,helmi_merger_2018}
or distinct chemodynamical sub-populations \citep[e.g.][]{kruijssen_formation_2019, horta_evidence_2020, buder_galah_2022},
which are the likely remnants of destroyed satellites that merged with the MW early in its formation. At the galaxy cluster
scale, \cite{deason_stellar_2021} showed that the stellar density profiles of simulated clusters
exhibit a well-defined edge, coincident with the `splashback' radius of the underlying DM halo
\citep[e.g.][]{adhikari_splashback_2014, diemer_splashback_2017}. This feature has since been detected observationally
\citep{gonzalez_discovery_2021}, indicating that the IHL may provide an observable probe of the underlying DM distribution.

Despite recent observational progress, accurately characterising the IHL for a representative sample of extragalactic galaxies
remains challenging. Surveys based on resolved stellar populations provide a wealth of information on halo stars, but these
measurements can only be obtained for nearby systems \citep[e.g.][]{barker_resolving_2009, ibata_large-scale_2013, harmsen_diverse_2017}.
Deep integrated light surveys provide an alternative approach to studying the IHL of extragalactic systems, where the surface
brightness profiles of galaxies can be decomposed based on assumptions about the analytic form appropriate for specific galactic components.
\cite{merritt_dragonfly_2016} measured the IHL of 8 nearby disc galaxies by decomposing their stellar mass surface density
profiles, finding an RMS scatter in the IHL mass fraction of approximately 1 dex, indicating that the IHL fraction varies
significantly, even for galaxies with similar morphologies and stellar masses. The IHL fractions of galaxy groups and clusters
at $z\approx0$ also exhibit significant scatter, with measured fractions ranging from a few per cent to $\approx 50$ per cent
\citep[e.g.][]{montes_faint_2022}. It is unclear how much of this scatter is inherently physical in origin
\citep[due to, for example, stochastic variations in formation histories, e.g.][]{fattahi_tale_2020, rey_how_2022}, and how much
is a result of the inconsistent methodologies adopted by different observational studies \citep[see, e.g.,][]{kluge_photometric_2021}.

The outlook for characterising the IHL consistently across different environments is perhaps more promising in cosmological,
hydrodynamical simulations, in which the various structural components of galaxies can in principle be identified from the
kinematics of their stellar particles. In particular, understanding the formation mechanisms of the IHL and how they vary as a
function of mass can be addressed with large volume cosmological simulations, where galaxies that form in a broad range of
environments can be studied and tracked over time \citep[e.g.][]{schaye_eagle_2015, pillepich_first_2018, dave_simba_2019}.

\cite{canas_stellar_2020} introduced an adaptive phase-space algorithm, specifically designed to distinguish kinematically hot stellar
particles (which they identified with the IHL) from centrally concentrated stellar structures and applied it to
galaxies identified in the Horizon-AGN simulation \citep{dubois_horizon-agn_2016}. This allowed them to study the mass-dependence
of the IHL fractions of a diverse population of galaxies, which revealed that the scatter in IHL masses is correlated with kinematic measures of galaxy
morphology. While their method was effective in identifying the IHL of galaxies in diverse environments, their results depend on a
free parameter that had to be calibrated to a small number of uncertain observations. 

Unsupervised machine learning techniques provide a viable alternative for identifying kinematically distinct
structures within simulated galaxies \citep[e.g.][]{domenech-moral_formation_2012}. \cite{obreja_nihao_2016} showed that Gaussian
Mixture Models (GMMs) can be applied to the stellar components of galaxies to identify discs, bulges, and stellar haloes, whose
properties resemble those of observed systems. Automating structural decomposition for a diverse galaxy sample is, however,
non-trivial. As a result, studies are typically limited to small samples of galaxies, for which the clusters of stars identified by
the GMMs can be manually assigned to different galactic components \citep{obreja_introducing_2018, obreja_nihao_2019}, or they are
limited to galaxies within a narrow range of mass or morphology (e.g. \citealt{ortega-martinez_milky_2022}, though see
\citealt{du_identifying_2019}).

In this work, we use GMMs to characterise the disc, bulge and IHL components of simulated galaxies across a wide range of halo masses.
We apply our methodology to galaxies from three simulations based on the \eagle\ model of galaxy formation, allowing us to study how
the mass fraction, stellar populations and structure of these components vary from the galactic scale up to the scale of massive galaxy
clusters. The remainder of this paper is organised as follows. In Section \ref{sec:sims}, we describe the \eagle\ simulations and
introduce the kinematic quantities used in our structural decomposition. In Section \ref{sec:decomp}, we introduce our decomposition
technique and compare our results to alternative techniques. In Section \ref{sec:results}, we analyse the stellar populations and
structural properties of the disc, bulge, and IHL components; and in Section \ref{sec:conc} we summarise our results.

\section{Simulations and analysis}\label{sec:sims}

\subsection{The \eagle\ simulations}

The \eagle\ Project \citep{schaye_eagle_2015, crain_eagle_2015} is a suite of cosmological, smoothed particle hydrodynamical
(SPH) simulations that model the formation and evolution of galaxies in a $\Lambda$CDM universe using cosmological parameters
consistent with the \citet{planck_collaboration_planck_2014} results. The majority of our analysis is based on the
${\rm L_{box}}=100$ cubic Mpc ``Reference'' run of the \eagle\ Project (i.e. Ref-L0100N1504; see Table 2 of
\citealt{schaye_eagle_2015}), which follows structure formation using ${\rm N_{DM}}=1504^3$ equal-mass dark matter (DM) particles
and initially the same number of baryonic particles. The mass of DM particles is $m_{\rm DM}=9.70\times 10^6\,{\rm M_\odot}$, and
$m_{\rm gas}=1.81\times 10^6\,{\rm M_\odot}$ is the initial baryonic particle mass.

Initial conditions were evolved to $z=0$ using an updated version of {\sc Gadget-3} \citep{springel_cosmological_2005,schaye_eagle_2015}
that includes subgrid models for, among other processes, radiative cooling and photoheating \citep{wiersma_chemical_2009},
star formation and stellar feedback \citep{schaye_relation_2008, dalla_vecchia_simulating_2012} the growth of supermassive black
holes (BHs) through mergers and accretion, and feedback from active galactic nuclei (AGN; \citealt{rosas-guevara_impact_2015}). The
subgrid model parameters were calibrated so that \eagle\ reproduced $z \approx 0$ observations of the galaxy stellar mass function,
the stellar size-mass relation, and the black hole mass-stellar mass relation \citep[see][for details]{crain_eagle_2015}. Subsequent work
has shown that \eagle\ also reproduces observations of galaxies at $z > 0$, such as their angular momenta \citep{lagos_angular_2017},
sizes \citep{furlong_size_2017}, velocity dispersion, and rotational velocities \citep{van_de_sande_sami_2019}, highlighting that galaxy
structure and kinematics are reproduced well by \eagle.

We supplement results from Ref-L0100N1504 at high and low halo masses using data from two other simulations.
One is the Cluster-\eagle\ project (C-\eagle; \citealt{bahe_hydrangea_2017, barnes_cluster-eagle_2017}), which is a suite of 30
resimulations of cluster-mass haloes; the other is a ${\rm L_{box}}=50\,{\rm Mpc}$ \eagle\ volume simulated using higher resolution in
the DM component while maintaining the original baryonic particle mass and force resolution that was used for \eagle\,
(see \citealt{ludlow_spurious_2023}, for details). We use the latter run, which is referred to in our paper as 50-HiResDM, to test the sensitivity of our results to the spurious collisional heating of stellar particles by DM halo particles \citep[see, e.g.,][]{ludlow_energy_2019, ludlow_spurious_2021, wilkinson_impact_2023}. 50-HiResDM employed the same subgrid models, as well as the same numerical and subgrid parameters as \eagle, but  its DM particle mass is
$m_{\rm DM}=1.39\times 10^6\,{\rm M_\odot}$, i.e. a factor of 7 lower than the value used for Ref-L0100N1504. The C-\eagle\, project used the
same subgrid and numerical set-up as \eagle, but adopted different parameters for the AGN feedback subgrid model to achieve better agreement
with the observed gas content of galaxy clusters (see \citealt{bahe_hydrangea_2017} and \citealt{barnes_cluster-eagle_2017} for details).

\subsection{Identifying DM haloes and galaxies}

DM haloes, their substructure haloes and associated galaxies were identified using \subfind\
\citep{springel_populating_2001, dolag_substructures_2009}. Haloes were first identified using a Friends-of-Friends algorithm
(FoF; \citealt{davis_evolution_1985}), which links nearby DM particles into FoF groups. Baryonic particles were assigned to the same group
as their nearest DM particle, provided it belonged to one. Each FoF halo was then divided into self-bound substructures, or ``subhaloes''
for short. One subhalo typically dominates the total mass of the FoF halo -- we refer to this as the ``central'' subhalo; lower-mass
subhaloes we refer to as ``satellite'' subhaloes. The baryonic particles associated with central and satellite subhaloes are
referred to as central and satellite galaxies, respectively. The stellar mass of a galaxy, $M_\star$, is defined as the total mass of all 
stellar particles gravitationally bound to a subhalo.

We restrict our analysis to stellar particles associated with central galaxies identified at $z=0$, but exclude those
bound to satellite galaxies. We note that this choice may be questionable for systems with large numbers of
satellite galaxies or for those undergoing mergers; in these cases, distinguishing stellar particles that are bound
to a central galaxy from those bound to its satellites is challenging. However, we find that MW-mass central
galaxies in \eagle\ typically contribute 97 per cent of the total stellar mass associated with their FoF haloes, with satellites
contributing $\lesssim 3$ per cent. At higher halo masses, however, where mergers and substructure are more prevalent,
the contribution of satellite galaxies to the total stellar mass of FoF haloes can be quite large, sometimes reaching as high
as $\approx 40$ per cent for halo masses $\gtrsim 10^{13}\,{\rm M_\odot}$. 
Satellite galaxies, however, typically dominate the stellar mass budget at larger galacto-centric radii than that which encompasses the
majority of the central's stellar material. For that reason, we neglect the possible contribution of satellites
to the IHL of galaxies.

We henceforth quantify halo masses using $M_{200}$, i.e. the total mass (DM plus baryonic) within the spherical radius 
$r_{200}$ that encloses a mean density of $200\times \rho_{\rm{crit}}(z)$, where $\rho_{\rm{crit}}(z)=3\,H(z)^2/8\pi G$ is
the critical density, $H(z)$ is the Hubble parameter, and $G$ is the gravitational constant.
The centres of haloes and galaxies are defined as the location of their DM particle with the lowest gravitational potential energy.

\subsection{Kinematic quantities used for structural decomposition}\label{subsec:analysis}
We begin by repositioning the stellar particles of each central galaxy relative to its halo centre. The velocity frame of the galaxy
is at rest with respect to the centre of mass motion of the innermost 80 per cent of its stellar mass. All positions and velocities
from herein refer to these recentred quantities. 

Galaxies are then oriented such that the $z$-axis aligns with their total stellar angular momentum vector, $\vec{J_\star}$, which is
calculated using all stellar particles between 2 and 30 kpc. The lower limit of 2 kpc is imposed to minimise contributions from stellar
particles with disordered motions, or from kinematically decoupled cores\footnote{Although these are rare in \eagle\ \citep{lagos_diverse_2022},
they can impact the measured stellar angular momentum significantly.}, while the upper limit minimises the contribution from particles that
do not belong to the central disc or spheroidal component of the galaxy. We have verified the orientation of $\vec{J_\star}$ is robust to reasonable variations to the upper and lower radial limits.

We use the following quantities to decompose \eagle\ galaxies into distinct structural components, which may include a disc, a bulge, and IHL:
\begin{itemize}

\item \jzjc: The specific angular momentum in the $z$-direction relative to the specific angular momentum of a particle on a circular orbit
  with the same binding energy. This quantity was first introduced by \cite{abadi_simulations_2003} to isolate disc stars;
  it is commonly referred to as the orbital circularity parameter. Particles with \jzjc\ values of 1 (-1) are on prograde (retrograde) circular
  orbits in the plane perpendicular to the net angular momentum vector (i.e., in the disc plane of a late type galaxy). 

\item \jpjc: The specific angular momentum in the plane parallel to $\vec{J_\star}$ (i.e., perpendicular to the disc of a late type galaxy),
  normalised by $j_{\rm{circ}}$ (note that in our coordinate system, $j_{\rm p}^2=j_x^2 + j_y^2$ ). This quantity was first introduced by
  \cite{domenech-moral_formation_2012} to aid in identifying disc stars.

\item \emin: The ratio of the specific binding energy of a particle to that of the most bound stellar particle in the galaxy.

\end{itemize}

For a given value of the specific binding energy, the maximum value of the specific angular momentum corresponds to that of a particle on
a prograde circular orbit, which we refer to as $j_{\rm{circ}}$ \citep{abadi_simulations_2003}. We estimate $j_{\rm{circ}}$ numerically using
the orbital information of stellar particles \citep[see also][]{thob_relationship_2019, kumar_galaxy_2021}. Particles are first divided into
150 equally-spaced bins of specific binding energy. Within each bin, the maximum value of the specific angular momentum of all particles, $j_{\rm max}$,
is taken to be the value of $j_{\rm{circ}}$ for particles within that bin. Due to the finite mass resolution of our simulations
and the diffuse nature of galaxy outskirts, this approach may be inaccurate at large galacto-centric distances, where bins naturally contain
fewer stellar particles than those in the central regions of galaxies. However, we have verified that the results of our galaxy decomposition
are insensitive to reasonable variations in the number of binding energy bins used to calculate $j_{\rm{circ}}$. This is because \jzjc\ is mostly useful for identifying disc particles, which are centrally concentrated and located in regions of a halo that are well sampled by stellar
particles.
\section{Identifying the structural components of simulated galaxies}\label{sec:decomp}

In this section, we introduce our structural decomposition technique and apply it to two
example galaxies identified in Ref-L0100N1504 (S\ref{subsec:decomp_expl}). In S\ref{subsec:vary_nc}, we test the effect of varying
the number of Gaussians used by the GMM on the resulting disc, bulge, and IHL mass fractions, and in S\ref{subsec:decomp_alt_methods}
we compare the results of our decomposition to alternative estimates of the kinematic morphologies and IHL fractions of galaxies. The results in this section are limited to central galaxies identified at $z=0$ in Ref-L0100N1504 that have $M_{200}> 10^{11.7}\,{\rm M}_\odot$,
resulting in a sample of 2415 galaxies. This halo mass limit ensures that the structural and kinematic properties of the galaxies are
robust to spurious heating by DM particles at their half stellar mass radii \citep[$r_{50}$; see Table 2 of][]{ludlow_spurious_2023}.

\subsection{Decomposing the stellar component of \eagle\ galaxies using GMMs}\label{subsec:decomp_expl}
\begin{figure*}
  \centering
  \includegraphics[width=\linewidth]{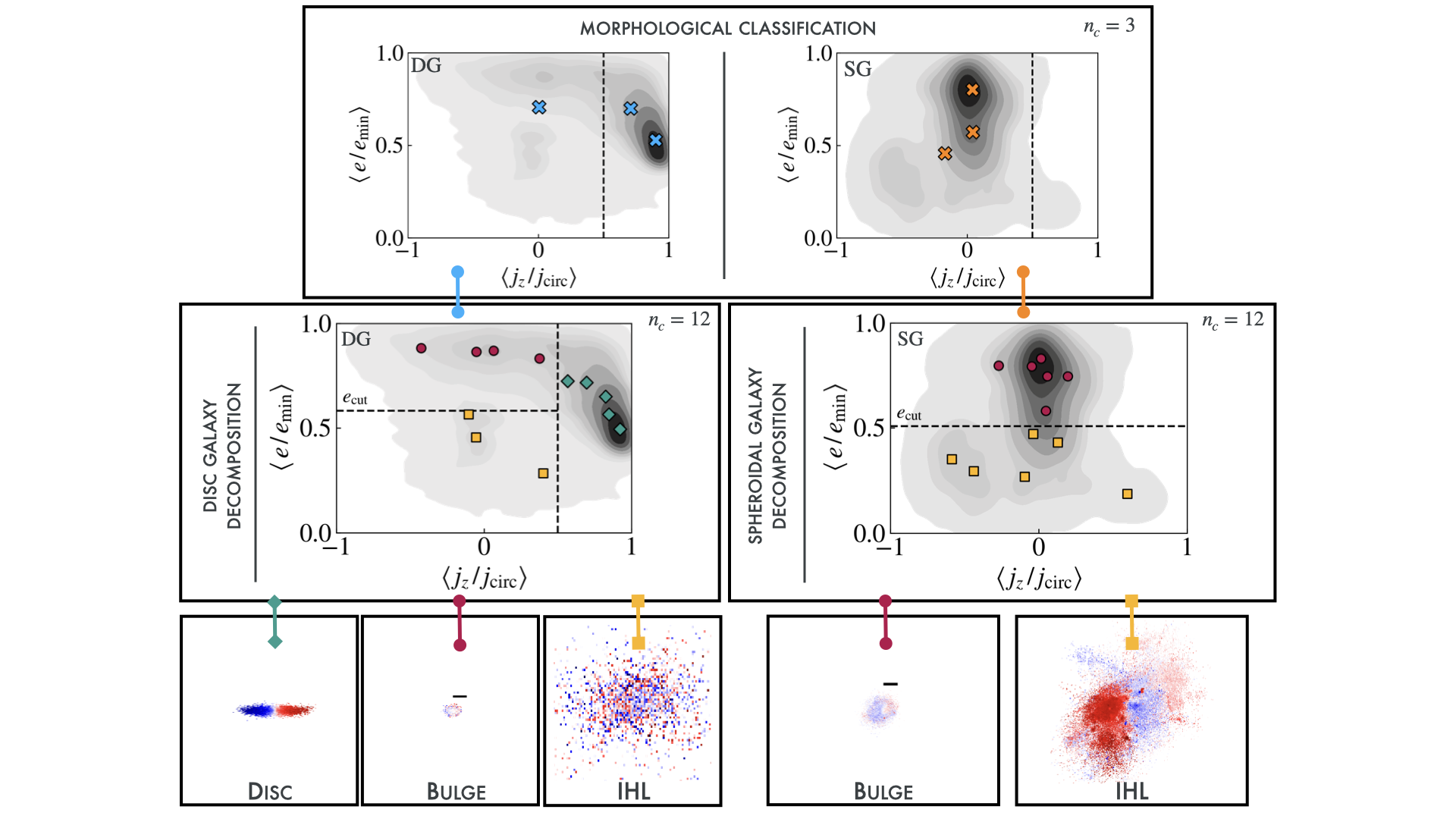}
  \caption{Visual summary of our decomposition technique using an example disc galaxy (DG; left panel) and spheroidal galaxy (SG; right panel). The top row depicts the initial morphological classification step and the middle row depicts the rules used to assign Gaussian components to galaxy structural components (i.e. disc, bulge, or IHL; see text for further details). The bottom row shows the resulting $x-z$ projections for each component, where cells have been colour-coded by the median value of $v_y$. The black lines in the "Bulge" panels indicate physical scales of 15 kpc and 250 kpc for DG and SG, respectively.}
  \label{fig:flowchart}
\end{figure*}

We use GMMs to decompose the stellar components of \eagle\ galaxies into at most three physical components -- a disc, a centrally
concentrated bulge, and an extended IHL -- but make no attempt to isolate dynamically distinct sub-components of them (i.e. we do not
distinguish between thin and thick discs, or classic- and pseudo-bulges). We follow \citet[][see also \citealt{du_identifying_2019}]{obreja_nihao_2016}
and use the kinematic parameter space of stellar particles defined by their values of \jzjc, \jpjc, and \emin. Our GMMs find particle clusters
in this parameter space and approximate their distributions with multi-dimensional Gaussians. The GMMs evaluate the probability of each stellar particle belonging to each of these Gaussian distributions. In this work, particles are wholly allocated to the Gaussian distribution to which they are assigned the largest probability (i.e. we adopt a hard classification) and these Gaussian components are later associated with
the different structural components of a galaxy.

The optimal number of Gaussian distributions, $n_c$, required to disentangle the various components of a galaxy is not known a priori,
so assumptions must be made. \citet{obreja_nihao_2019} applied GMMs to 25 galaxies simulated at high resolution using $n_c\leq 5$ and associated
each of the best-fit Gaussians to one physical galaxy component: a thin or thick disc, classical- or pseudo-bulge, or a stellar
halo. However, the structural components of galaxies may not follow simple Gaussian distributions in the input parameter space (e.g. they may
posses kinematic substructure or exhibit non-Gaussian distribution functions), and when they do not multiple Gaussian distributions per galaxy
component seem to fare better. \citet{du_identifying_2019} used a modified Bayesian information criterion to determine that
$5\lesssim n_c \lesssim 12$ works well for galaxies in Illustris-TNG100 \citep{pillepich_first_2018} with stellar masses
$M_\star \gtrsim 10^{10}\, {\rm M_\odot}$; they assigned each Gaussian to a physical galaxy component based on the values\footnote{We follow the
nomenclature of \cite{du_identifying_2019} and denote the mean of the Gaussians output by our GMM fits using angular brackets, e.g. \meanjzjc\
or \meanemin.} of \meanjzjc and \meanemin.

\subsubsection*{Morphological classification}
We follow a different approach, and initially set $n_c=3$ in order to assess whether each galaxy possesses a significant disc component or
if it can be approximated as a pure spheroid. If one or more of the best-fit Gaussians have a mean circularity \meanjzjc $\geq 0.5$
we classify the galaxy as a disc (and hereafter refer to them as ``disc'' galaxies), otherwise it is spheroid dominated (hereafter, ``spheroid''). 
After some experimentation, we found that this initial morphological classification prevents the identification of spurious discs in
dispersion dominated galaxies for GMMs run with larger $n_c$, cases that often lead to net counter-rotating
spheroidal components. This occurs because dispersion supported systems often contain a small but significant
fraction of stellar orbits with high \jzjc\ values, and assigning those orbits to a ``disc'' skews the
distribution of the remaining \jzjc\ values lower. The top panels of Fig.~\ref{fig:flowchart} show two typical
galaxies that were classified as a disc (left panel) and spheroid (right panel) using $n_c=3$. 

After categorising galaxy morphologies this way, we again run GMMs but this time using a range of $n_c$ values. For the bulk
of our analysis we adopt $n_c=12$, but discuss below how increasing or decreasing $n_c$ affects the median mass fractions of the
various structural components of galaxies inferred from GMMs, as well as the intrinsic variation in them for individual galaxies.

\subsubsection*{Disc galaxy decomposition}
For all $n_c$, disc galaxies are modelled using the (\jzjc, \jpjc, \emin) parameter space of stellar particles and we assign the corresponding best-fit Gaussian distributions to one of the three physical galaxy components based on their values of \meanjzjc and \meanemin.
Those with \meanjzjc $\geq0.5$ are assigned to the disc, and the rest are split between the bulge and IHL. To do so, we identify the
Gaussian distributions with the maximum and minimum \meanemin\ values and define $e_{\mathrm{cut}}$ as the midpoint between them, i.e.
$e_{\mathrm{cut}}=[\min(\langle e/e_{\rm min}\rangle) + \max(\langle e/e_{\rm min}\rangle)]/2$. The remaining Gaussian clusters not
assigned to the disc (i.e. those with \meanjzjc$< 0.5$) are assigned to the bulge if \meanemin$\geq e_{\mathrm{cut}}$, or to the IHL
if \meanemin$<e_{\mathrm{cut}}$. Note that there is the chance that the IHL will not be detected, as is the case for the $n_c=3$
disc model in Fig. \ref{fig:flowchart} (upper-left panel). This occurs most often when $n_c$ is small and the IHL is significantly
less massive than the disc and bulge components.

\subsubsection*{Spheroidal galaxy decomposition}
We run the same GMMs for spheroids, but this time using (\jzjc, \emin) as the input parameter space (the broad \jpjc\ distributions
for the bulges and IHL of spheroidal galaxies overlap considerably making this quantity less useful for distinguishing these components). As for discs, Gaussians with \meanemin$\geq e_{\mathrm{cut}}$ are assigned to the
bulge; those with \meanemin$<e_{\mathrm{cut}}$ are assigned to the IHL\footnote{If we instead adopt a fixed specific binding energy cut of $e/e_{{\rm min}}=0.5$ to distinguish the bulge and IHL components, bulge masses are systematically higher by about 5 per cent; IHL masses are systematically lower by roughly 20 per cent.}.

\subsubsection*{Applying the method to example galaxies}
The middle row of Fig. \ref{fig:flowchart} show the results of running a GMM using $n_c=12$ on the disc and spheroidal galaxies
mentioned above, and in the bottom row we plot $x-z$ projections of the stellar particles assigned to each component (these galaxies are hereafter referred to as DG and SG, respectively). DG (left) is roughly the mass of the MW, having 
$M_\star \approx 10^{10.7}\,\rm{M_\odot}$ and $M_{200}\approx 10^{12.3}\,M_\odot$. SG is the central galaxy of a low-mass cluster with $M_\star\approx 10^{12.1}\,\rm{M_\odot}$ and $M_{200}=10^{14.3}\,{\rm M_\odot}$. Teal diamonds, maroon circles, and yellow squares show the values of \meanemin\ and \meanjzjc\ corresponding to the best-fit Gaussians assigned to the disc, bulge, and IHL, respectively.

The vertical black lines show the value of \meanjzjc$=0.5$ used to distinguish Gaussians assumed to
represent the disc component from those that represent the bulge or IHL. The horizontal black lines show the values of
$e_{\rm{cut}}$. For DG, we find $e_{\rm{cut}} = 0.59$ whereas for SG
$e_{\rm{cut}} = 0.50$. We note that these values are lower than the value of 0.75 adopted in previous work
\citep[e.g.][]{du_kinematic_2020, du_evolutionary_2021} to distinguish between GMM Gaussians representing the bulges and IHL of simulated
disc galaxies, but the most appropriate value of $e_{\rm cut}$ for a particular galaxy is unclear.
Recent work suggests that the optimal cut in binding energy may depend on galaxy morphology \citep[e.g.][]{zana_morphological_2022}.
Nonetheless, when applied to DG, our method effectively separates the tightly-bound stellar structures (i.e. the bulge) from the
loosely-bound ones  (i.e. the IHL). For SG, the separation between the bulge and IHL components is less
clear: several Gaussians have \meanemin$\approx e_{\rm{cut}}$, suggesting that the bulge and IHL components of this galaxy
are less dynamically distinct than those of DG.

\begin{figure}
  \centering
  \includegraphics{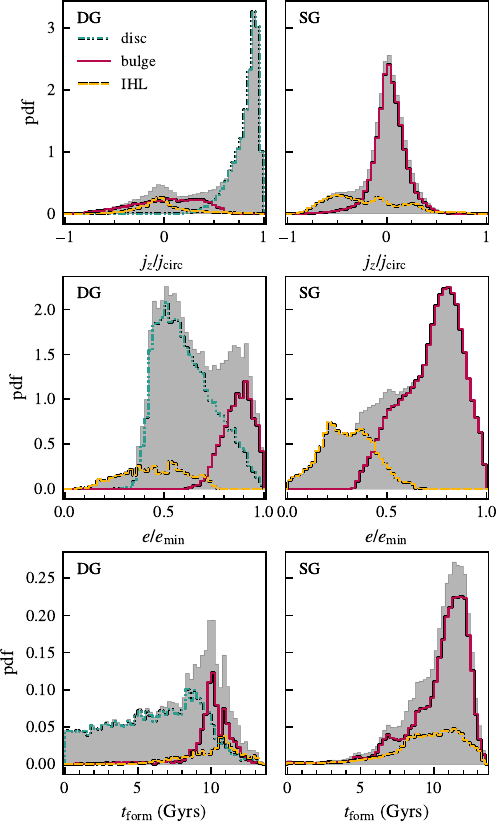}
  \caption{From top to bottom, various rows show the \jzjc, \emin, and $t_{\rm form}$ distributions, respectively, for the example disc (DG; left panels) 
    and spheroidal galaxy (SG; right panels). The distributions for all stellar particles associated with the galaxies are indicated by the grey shaded histograms;
    the dash-dotted teal lines, solid maroon lines, and dashed yellow lines show the distributions for stellar particles assigned to the disc, bulge, and IHL
    components.}
  \label{fig:cs_props}
\end{figure}

Fig. \ref{fig:cs_props} shows a few properties of the stellar particles belonging to the different structural
components of DG (left column) and SG (right column;
in both cases, the structural components were identified using a GMM with $n_c=12$). The top and middle
rows show the distributions of \jzjc\ and \emin, respectively; the bottom row shows the distributions of stellar ages ($t_{{\rm form}}$).
Grey histograms in each panel represent all stellar particles belonging to each galaxy, and the colored lines
show the subset of stellar particles assigned to the disc (teal dotted-dashed lines), bulge (solid maroon lines),
and IHL (dashed yellow lines). 

DG is composed of three distinct components: a rotationally supported disc, a tightly bound bulge and a loosely bound
stellar halo (the latter two components are largely dispersion supported, as expected). The bottom left panel of Fig.
\ref{fig:cs_props} shows that the disc component formed over an extended time period, with star formation peaking $\approx 9$
Gyrs ago, and gradually tapering off to roughly half of the peak value by $z=0$. The disc, which is composed primarily of stellar
particles that formed in-situ ($f_{\rm{ex-situ}}\approx 0.1$),\footnote{We use the ex-situ classification of \cite{davison_eagles_2020}.
Stellar particles are traced back to the snapshot prior to star formation: if the associated gas particle does not belong to the main branch of the $z=0$ subhalo at this snapshot, the stellar particles are flagged as having formed ``ex-situ''. All other particles are flagged as having formed ``in-situ''.} has a half mass stellar age of 6.4 Gyr, and an interquartile age range of
about 5 Gyr.

The bulge component of DG contains a slightly higher contribution from ex-situ stars ($f_{\rm{ex-situ}}\approx 0.18$) and is, on average,
composed of the oldest stellar populations in the galaxy; its half mass stellar age is $t_{50} \approx 10.1\, {\rm Gyrs}$. The IHL component is the
most extended in the galaxy (its half stellar mass radius is $r_{50}=30.9\,{\rm kpc}$; for the bulge, $r_{50}=2.8\,{\rm kpc}$) and is
dominated by stars that formed ex-situ ($f_{\rm{ex-situ}}=0.73$), suggestive of a merger-driven formation scenario. Similar to the bulge
component, the IHL hosts a relatively old stellar population with a half mass age of about 10.1 Gyrs. Neither the bulge nor the IHL contain an
appreciable number of stellar particles with ages $\lesssim 6\,{\rm Gyr}$ (the disc formed 54 per cent of its stellar mass in that time).

The bulge component of SG is similar to that of DG: it is dispersion-supported and comprised primarily of old stellar populations (the
half mass age of bulge stars in SG is $\approx 11.0\, {\rm Gyrs}$). The visible peaks in the $t_{{\rm form}}$ distribution of bulge stars
does, however, indicate that the bulge contains several distinct stellar populations. Together with the high ex-situ fraction
($f_{\rm{ex-situ}}\approx 0.80$), this implies a formation history dominated by multiple merger events, consistent with the standard
model for the formation of brightest cluster galaxies (BCGs) in $\Lambda{\rm CDM}$
\citep[e.g.][]{de_lucia_hierarchical_2007, robotham_galaxy_2014}. The IHL of SG is also dominated by ex-situ stars
($f_{\rm{ex-situ}}\approx 0.94$) and is comprised of two kinematically-distinct components: one dispersion-supported structure with a
peak orbital circularity of \jzjc$\approx 0$, and another with a peak at \jzjc $\approx -0.5$. The latter component counter-rotates with
respect to the net angular momentum of the galaxy and is visible in the plot of the component projections (see bottom panel of Fig \ref{fig:flowchart}).

\begin{figure}
  \centering
  \includegraphics{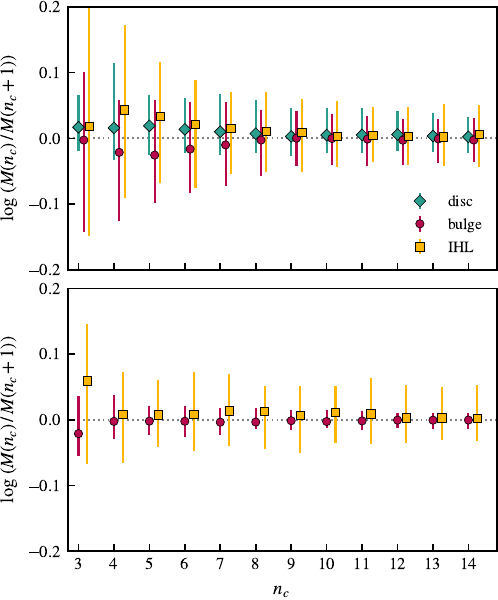}
  \caption{Logarithmic difference in the estimated mass of each structural component induced by increasing $n_c$ by 1, plotted as a function of $n_c$ (for visual clarity, points corresponding to the disc and IHL components have been offset slightly from the integer $n_c$ values). Medians are plotted in diamonds, circles, and squares for the disc, bulge and IHL components, respectively; error bars represent the interquartile range. Results for disc galaxies are plotted in the top panel; results for spheroidal galaxies are show in the bottom panel.}
  \label{fig:mass_by_n}
\end{figure}

\subsection{The impact of varying $n_c$ on the decomposition results}\label{subsec:vary_nc}

Fig. \ref{fig:mass_by_n} shows the logarithmic change in the stellar mass that is allocated to the disc
(teal diamonds; upper panel), bulge (maroon circles), and IHL (yellow squares) components of individual galaxies as
$n_c$ is increased by $\Delta n_c=1$, starting from $n_c=3$ and increasing to $n_c=14$. The top and bottom panels show results separately for
the subset of disc and spheroidal galaxies, respectively. Symbols correspond to the median\footnote{When calculating
this ratio, we set $\log({\rm M}(n_c) / {\rm M}(n_c+1) ) = -1$ if a particular component is undetected for a given value of
$n_c$. Doing so allows us to include such instances in our estimates of the scatter and logarithmic change in component masses
as $n_c$ is increased, which would otherwise be biased by only including systems for which masses can be estimated. We note
that such occurrences do not affect the medians values plotted in Fig.~\ref{fig:mass_by_n}.} values of $\log[M(n_c) / M(n_c+1)]$,
and error bars show in interquartile scatter (note: $M(n_c)$ is used generically here to represent the mass assigned to a
particular galaxy component after running a GMM that uses $n_c$ Gaussians).

The top panel of Fig. \ref{fig:mass_by_n} shows that, for disc galaxies, the mass assigned to each
structural component is typically robust provided $n_c\geq 8$. 
For spheroids, the situation is similar. The median masses assigned to the bulge and IHL of individual
systems is largely unchanged when $n_c$ is increased from $\approx 4$, although the scatter in
mass assigned to the IHL of individual galaxies is larger than that of bulges for all $n_c$. The larger scatter in the mass assigned the IHL component compared to the bulge of spheroidal galaxies is due its relatively low mass, 
which makes it more susceptible to small changes in $M(n_c)$. Note, however, that both the median mass fraction and
scatter in mass assigned to each
component, including the IHL, are well-behaved provided $n_c\geq 9$, regardless of galaxy morphology. 

We adopt $n_c=12$ for our analysis for the following reason. If the Gaussian distributions are equally divided between
the structural components of galaxies, it implies that the disc, bulge, and IHL of disc-dominated galaxies will
each be identified by 4 Gaussian distributions; for spheroidal galaxies, the bulge and IHL will be identified by
6 Gaussians. The exact division of the Gaussians among the structural components of galaxies will, of course,
vary from galaxy to galaxy, but allowing for multiple Gaussians per galactic component can better accommodate
the presence of distinct sub-populations.

\begin{figure}
  \centering
  \includegraphics{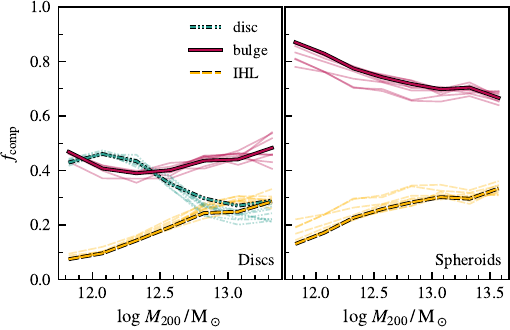}
  \caption{The median stellar mass fraction assigned to each galactic component as a function of $M_{200}$. The dash-dotted teal lines, solid maroon lines, and dashed yellow lines show the median mass fractions of the disc, bulge, and IHL components, respectively. The thick lines show results obtained using $n_c=12$; results for other $n_c\geq8$ are plotted as faint lines of the corresponding colour and style. Results are shown separately for discs (left panel), and spheroids (right panel).}
  \label{fig:fcomp}
\end{figure}

We stress, however, that the exact value of $n_c$ is somewhat arbitrary: provided 
$8\lesssim n_c \lesssim 15$, the masses allocated to the different structural components of most galaxies
are relatively stable. This result is shown another way in Fig.~\ref{fig:fcomp}, where
we plot the fraction of mass assigned to the various structural components of discs (left) and spheroids
(right) as a function of $M_{200}$. The thick lines in each panel correspond to results obtained for $n_c=12$;
the thin lines show results for other values of $n_c$ in the range $8\leq n_c \leq 15$. Note that the variation in the component mass fractions that arise from changing $n_c$ is smaller than the differences between the mass fractions of each component.

\subsection{Comparison to alternative definitions of galaxy components}\label{subsec:decomp_alt_methods}

\subsubsection{Relation to kinematic estimates of galactic morphology}
\begin{figure}
  \centering
  \includegraphics[width=\linewidth]{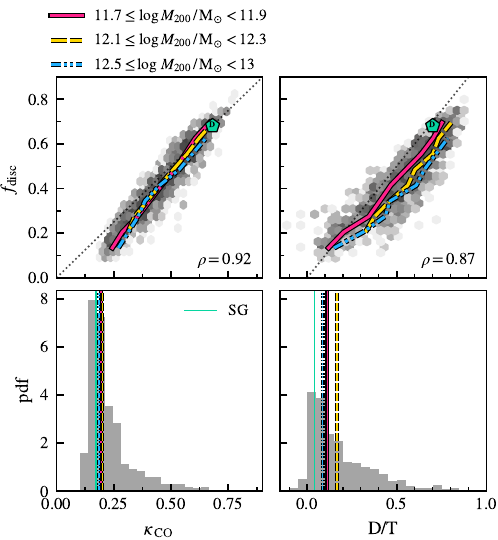}
  \caption{Top row: The \fdisk-\kappaco\ (left panel) and \fdisk-D/T (right panel) distribution for discs. Median
    relations are plotted for 3 bins of halo mass, as indicated above the plot. The black dotted line shows the one-to-one line for reference and our example disc galaxy (DG) is plotted using an outsized green pentagon. The Spearman correlation coefficient ($\rho$) for the total disc sample is displayed in the bottom right corner of the upper panels. Bottom row: \kappaco\ and D/T distributions for spheroids. Vertical lines indicate the median values for the same three halo mass bins and the vertical green line shows values for our example spheroidal galaxy (SG).}
  \label{fig:fdisc_comp}
\end{figure}

In Fig. \ref{fig:fdisc_comp} we compare the mass fraction allocated to the disc component by our GMM (\fdisk) with two other
kinematic indicators\footnote{Specifically, we define $\kappa_{\rm co}=(2\,K_\star)^{-1}\sum_{j_{z,k}>0} m_k\,(j_{z,k}/R_k)^2$, where $K_\star$ is the total kinetic energy of the stellar particles, $j_{z,k}$ is the $z$-component of the angular momentum of particle $k$, and $R_k$ is its distance from the $z$-axis. The disc-to-total ratio is defined as $D/T=1-S/T=1-2/M_\star \sum_{j_{z,k}<0} m_k$, where $S/T$ is the spheroid-to-total ratio and $m_k$ is the mass of the $k^{\rm th}$ stellar particle.} of galaxy morphology: the fraction of kinetic energy
in co-rotation (\kappaco; see \citealt{correa_relation_2017}) and the disc-to-total ratio (D/T; see e.g.
\citealt{thob_relationship_2019}). Both quantities were
measured using stellar particles that lie within a spherical 30 kpc aperture. 

The top panels of Fig. \ref{fig:fdisc_comp} plot the 2D histogram of \fdisk\ versus \kappaco\ (left) and D/T (right) for
disc galaxies. The lines show the median relations separately for three bins of $M_{200}$. The Spearman correlation
coefficient for the full sample of discs (labelled $\rho$ in the upper panels) is displayed in the bottom right corner of each panel.

Regardless of halo mass, there is a strong correlation between \fdisk\ and \kappaco\,  ($\rho=0.92$ for
all galaxies, and $\rho>0.89$ for the individual mass bins) that closely follows the one-to-one line. Close inspection,
however, reveals that the relation has a slope that is slightly steeper than 1: for \fdisk$\,\lesssim 0.4$
there is a tendency for \kappaco\, to exceed \fdisk. This is because our GMMs can in principle yield $f_{\rm disc}=0$,
whereas the minimum value of \kappaco\, for isotropic, dispersion supported systems with no rotation is
$\kappa_{\rm co}\approx 0.17$. This naturally biases the relation between \fdisk\, and \kappaco, particularly
for galaxies with small disc fractions.

The top right panel of Fig. \ref{fig:fdisc_comp} shows that \fdisk\ and D/T are also
strongly correlated ($\rho=0.87$), but that D/T is slightly higher than \fdisk\, for the majority of galaxies. 
This is true for all mass bins, but a larger offset is seen for higher $M_{200}$. This offset occurs
because the D/T statistic implicitly assumes that all spheroidal components of galaxies do not rotate and are
completely dispersion supported, attributing all net rotation in the galaxy to the disc. A similar offset between
D/T and disc mass fractions obtained from GMMs was reported by \cite{obreja_nihao_2016}. The mass dependence of the
offset hints at an increasing prevalence of rotational support in the spheroidal component for galaxies in higher
mass haloes. 

The bottom panels of Fig. \ref{fig:fdisc_comp} show the distributions of \kappaco\ and D/T for  our sample of spheroids. The vertical pink, yellow, and blue lines show the median values for each mass bin and the green 
line shows the values obtained for the spheroidal galaxy used for Figs. \ref{fig:flowchart} and \ref{fig:cs_props} (i.e. SG).
The median \kappaco\ value for all spheroids is 0.19, only slightly larger than the value expected for
isotropic, dispersion supported systems with no net rotation. There are a handful of
spheroids with relatively high values of \kappaco\ and D/T (for example, 6.6 per cent of spheroids
have \kappaco$\,\ge 0.4$, whose mean D/T ratio is 0.6). A visual inspection of these galaxies
reveal they typically have lenticular morphologies, consistent with fast rotators \citep[e.g.][]{cappellari_structure_2016}, or
have experienced recent mergers, which complicates the disc-bulge decomposition.

The strong correlations between \fdisk\, and these alternative morphology metrics are perhaps unsurprising given that both \kappaco\,
and D/T are calculated directly from the $z$-component of the angular momentum.  Although
they are not completely independent measures of galactic morphology, the close correspondence between them indicates that our galaxy
decomposition technique yields sensible results and that the kinematic morphologies we recover are in agreement with previous work. 

\subsubsection{Comparison to alternative definitions of the IHL}
\begin{figure}
  \centering
  \includegraphics[width=\linewidth]{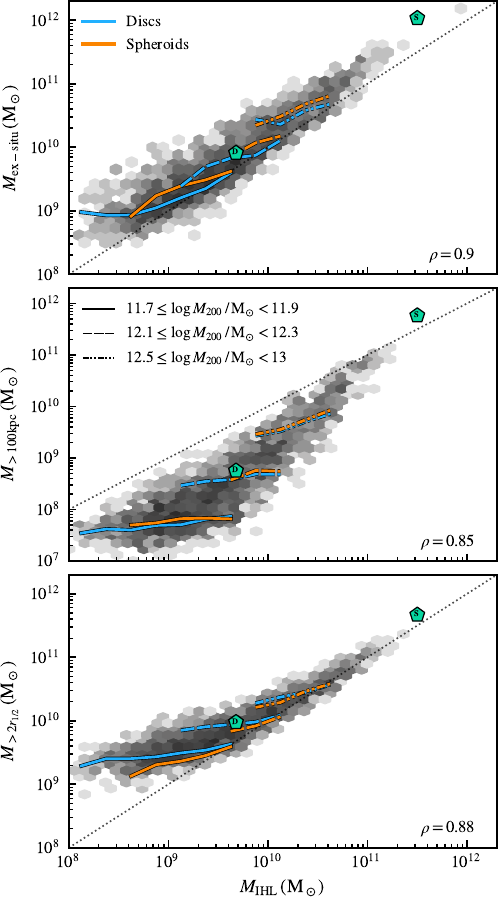}
  \caption{Correlations between various estimates of IHL mass. From top to bottom, respectively, the panels plot $M_{{\rm ex-situ}}$ (i.e. the total stellar mass that formed ex-situ), $M_{>100\,{\rm kpc}}$ (i.e. the stellar mass at $r > 100\,{\rm kpc}$), and $M_{>2\,r_{50}}$ (i.e. the stellar mass beyond two stellar half mass radii) versus $M_{\rm{IHL}}$. Median relations are plotted for discs (blue) and spheroids (orange) separately, in 4 bins of halo mass. The one-to-one line is shown as a dotted black line for reference. The locations of our example disc and spheroidal galaxies are plotted using green pentagons (labelled "D" and "S", respectively).}
  \label{fig:fihl_comp}
\end{figure}

We next compare the estimates of $M_{\rm{IHL}}$ obtained from our GMMs
to those obtained using three alternative IHL definitions: 1) the total stellar mass that formed ex-situ ($M_{\rm{ex-situ}}$; e.g. \citealt{cooper_galactic_2010}); 2) the total stellar mass at
$r>100\,\rm{kpc}$ ($M_{>100\,\rm{kpc\,}}$; e.g. \citealt{pillepich_halo_2014}); and 3) the total stellar mass beyond 2 times
the stellar half-mass radius of a galaxy ($M_{>2\,r_{50}}$; e.g. \citealt{elias_stellar_2018}).

The top panel of Fig. \ref{fig:fihl_comp} shows the relationship between $M_{\rm{IHL}}$ and $M_{\rm{ex-situ}}$. Median relations
are shown separately for discs and spheroids (blue and orange lines, respectively) in four bins of $M_{200}$
(see the legend in the middle panel). Both $M_{\rm{IHL}}$ and $M_{\rm{ex-situ\,}}$ increase with increasing
$M_{200}$, as seen by the separation of lines of different type (they move up and to the right as mass increases). 
As a result, the values of $M_{\rm{IHL}}$ and $M_{\rm{ex-situ\,}}$ for the whole sample of galaxies
are strongly correlated ($\rho=0.9$), although at fixed halo mass the correlations are somewhat weaker (the Spearman rank coefficients for
the various mass bins plotted in Fig.~\ref{fig:fihl_comp} range from 0.57 to 0.79).

$M_{\rm{IHL}}$ is typically larger than $M_{\rm{ex-situ}}$ by about 0.2 dex, although this offset increases with increasing $M_{200}$. The offset between $M_{\rm{IHL}}$ and $M_{\rm{ex-situ\,}}$ among massive spheroids likely reflects the fact that both the bulge and IHL components of these galaxies are dominated by stars that formed ex-situ. Associating the IHL
exclusively with accreted stellar material is therefore inappropriate for these systems, because much of the bulge mass also formed ex-situ \citep[e.g.][]{pillepich_halo_2014}.
We will return to these points
in the next section.

Given its extended nature, defining the IHL with an aperture-based approach is common; in this case, all mass beyond some radius is assigned to the IHL, whereas the mass within that radius is assumed to belong to the other galaxy components. The middle and bottom panels of Fig.~\ref{fig:fihl_comp} show the relation between $M_{\rm IHL}$ and two aperture-based IHL mass measurements: $M_{>100\rm{kpc}}$ \citep[used in e.g.][]{pillepich_first_2018}, and  $M_{>2\,r_{50}}$ \citep[used in e.g.][]{elias_stellar_2018},
respectively.

We find that $M_{>100\rm{kpc\,}}$ is lower than $M_{\rm IHL}$ for $\approx$ 99 per cent of galaxies (not surprisingly, only the most massive galaxies have $M_{>100\rm{kpc\,}} > M_{\rm IHL}$).
The two IHL mass estimates are correlated for the whole population ($\rho=0.85$) but at fixed $M_{200}$ they are less correlated ($\rho >0.6$ only for the highest halo mass bins plotted). This
suggests that the correlation between these IHL mass estimates is primarily driven by the fact that $M_{>100\rm{kpc\,}}$
and $M_{\rm IHL}$ both increase with increasing $M_{200}$. A similar conclusion applies to the relationship between $M_{>2\,r_{50}}$ and
$M_{\rm IHL}$. Note that $M_{>2\,r_{50}}$ is typically larger than $M_{\rm IHL}$, and has a much smaller halo-to-halo scatter
\citep[see][for a similar finding]{canas_stellar_2020}.

Although the IHL masses estimated using these alternative methods correlate with our measurements, it is clear that none of them
reproduce our results, and fare even poorer when comparisons are made at fixed halo mass. This is due to the fact that the
various components of galaxies do not have well-defined edges, nor are they composed purely of in-situ or ex-situ stars.
\section{Results}\label{sec:results}

In this section, we supplement the Ref-L0100N1504 sample with 897 additional galaxies from
50-HiResDM,\footnote{All galaxies in 50-HiResDM have virial masses $M_{200}\geq 10^{11.2}\,{\rm M_\odot}$, and are expected
to be robust to spurious collisional effects at their stellar half mass radii \citep[see][for details]{ludlow_spurious_2023}.}
as well as the 30 galaxy clusters from the C-\eagle~ suite.
This provides us with a sample of 3342 galaxies spanning the mass range 
$10^{11.2}{\rm M}_\odot\lesssim M_{200} \lesssim 10^{15.4}\,{\rm M}_\odot$ (or 
$10^{8.4} {\rm M}_\odot \lesssim M_\star\lesssim 10^{12.9}\,{\rm M}_\odot$ in stellar mass).
Of the full galaxy sample, $\approx 71$ per cent were classified as discs based on our morphological classification step
(see Section~\ref{subsec:decomp_expl}); the remainder as spheroids.
Only 15 galaxies have no discernible IHL component; these are primarily low-mass, disc-dominated systems (with median $M_\star=5.2\times10^9\,\rm{M_\odot}$, which corresponds to the mass of $\approx$ 3000 primordial gas particles). Higher baryon mass resolution is required to meaningfully study the IHL of these systems.
All results that follow were obtained from our GMM galaxy decomposition using $n_c=12$.

\subsection{The stellar-to-halo mass relation and morphology}\label{subsec:shmr}

\begin{figure}
    \centering
    \includegraphics{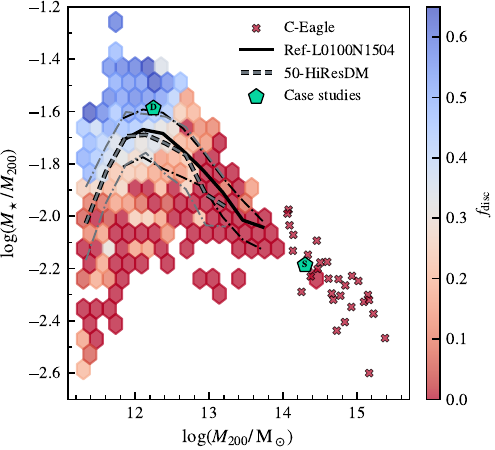}
    \caption{The stellar-to-halo mass relation for our full galaxy sample. Individual galaxies have been binned and plotted as hexagons that have been color-coded by their median value of $f_{\rm disc}$. The median relation for Ref-L0100N1504 (50-HiResDM) is plotted using a solid black (dashed grey) line; the 25$^{\rm th}$ and 75$^{\rm th}$ percentiles are plotted using dash-dotted lines. The two example galaxies discussed in Section~\protect\ref{sec:decomp} are highlighted using green pentagons. Individual C-\eagle~ galaxies are plotted as crosses. Note that our dynamical decomposition of galaxy components implies a strong morphology dependence to the stellar-to-halo mass relation. }
    \label{fig:fstar_morph}
\end{figure}

In Fig. \ref{fig:fstar_morph} we plot the stellar-to-halo mass relation (SHMR) for the full galaxy sample, colour-coded by \fdisk. The median relations obtained from Ref-L0100N1504 (shown as a solid black line) and 50-HiResDM (thick dashed line) are in good agreement \citep[see also][]{ludlow_spurious_2023}. Fig. \ref{fig:fstar_morph} also shows that, over the mass range $11.5 \lesssim \log\,(M_{200}/ {\rm M}_\odot) \lesssim 12.5$, there is a clear relationship between the stellar mass fraction of a galaxy and \fdisk: at fixed halo mass, disc galaxies have higher stellar masses than spheroids. For example, galaxies with $f_{\rm disc}>0.5$ have, on average, $M_\star\approx 2.1\times 10^{10}\,{\rm M_\odot}$; those with $f_{\rm disc}<0.1$ have $M_\star\approx 1.4\times 10^{10}\,{\rm M_\odot}$.
These results provide additional dynamical evidence for a morphology-dependent stellar-to-halo mass relation, consistent with the observational results of \cite{posti_dynamical_2021}, and the theoretical results of
\cite{correa_dependence_2020}, who reported a similar trend for \eagle\ galaxies but using \kappaco\ as a proxy for morphology.

Fig. \ref{fig:fstar_morph} also demonstrates that disc-dominated galaxies tend to occupy haloes with masses $M_{200}\lesssim 10^{12.5}\,{\rm M_\odot}$ (including 95 per cent of galaxies with \fdisk$> 0.5$), but above this mass spheroids dominate (e.g. at $M_{200}\gtrsim 10^{12.5}\,{\rm M_\odot}$, more than half of the galaxies in our sample have \fdisk$=0$), a trend that is also well-established observationally \citep[e.g.][]{posti_dynamical_2021}.

\subsection{The SFRs, ex-situ fractions, and ages of discs, bulges and the IHL}\label{subsec:scaling}

\begin{figure}
    \centering
    \includegraphics[width=0.98\columnwidth]{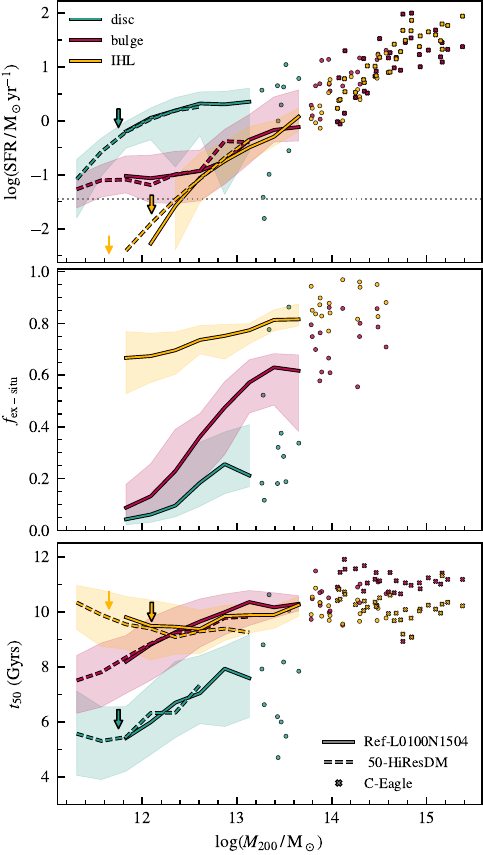}
    \caption{Different panels, from top to bottom, plot the median star formation rate (SFR), ex-situ fraction ($f_{\rm{ex-situ}}$), and stellar half-mass formation time ($t_{50}$) as a function of $M_{200}$. Results are shown separately for the disc (teal lines), bulge (maroon lines), and IHL components (yellow lines). The solid (dashed) lines indicate the median relations from Ref-L0100N1504 (50-HiResDM); shaded regions enclose the interquartile range. Medians are displayed only for mass bins that contain at least 10 galaxies, otherwise individual galaxies are plotted as circles (for Ref-L0100N1504) or crosses (for C-\eagle). The arrows with (without) black boundaries indicate the halo mass above which each component is resolved by more than 1000 (100) stellar particles for at least 90 per cent of galaxies. The dotted horizontal line in the top panel indicates the point at which we consider the SFR to be poorly resolved, which corresponds to a SFR of 10$\,m_{\rm{gas}}/500\,\rm{Myr}$.}
    \label{fig:meds}
\end{figure}

Fig. \ref{fig:meds} plots the median star formation rates (SFRs), ex-situ fractions ($f_{{\rm ex-situ}}$), and stellar half-mass formation times ($t_{50}$), versus $M_{200}$. Results are shown separately for each galaxy component and for all simulations (Ref-L0100N1504 results are shown using solid lines; 50-HiResDM results using dashed lines; and C-\eagle~ results using crosses). The SFRs were averaged over a lookback time of 500$\, \rm{Myr}$, but we have verified that our results are qualitatively insensitive to reasonable variations in that timescale. Note that merger trees are not available for our 50-HiResDM run, nor for C-\eagle, so we only present ex-situ fractions for Ref-L0100N1504. Also note that the limited baryonic mass resolution of our simulations, compounded with low component mass fractions, can result in galaxy components being resolved by very few stellar particles, particularly for galaxies occupying low-mass haloes. For that reason, in Figs. \ref{fig:meds} and \ref{fig:r50} we highlight the halo mass above which the various components are resolved by at least $1000$ (arrows with black boundaries) or $100$ (coloured arrows) stellar particles and for 90 per cent of galaxies. Note that the bulge components of galaxies are always resolved by at least 1000 stellar particles, and the discs by at least 100. We have also verified that the qualitative differences between the disc, bulge, and IHL populations discussed in this section are unchanged if we restrict our sample to components with at least 1000 particles.

Overall, the results plotted in Fig.~\ref{fig:meds} are in qualitative agreement with observational expectations and theoretical results: compared to bulges and the IHL, discs have the highest SFRs, the lowest fractions of ex-situ stars, and are typically the youngest component of a galaxy \citep[see, e.g.,][for observational evidence of the latter]{robotham_profuse_2022}. Note that discs have small but non-zero ex-situ fractions, possibly due to the presence of stellar particles tidally stripped from satellites co-planar with the disc \citep[e.g.][]{abadi_simulations_2003}.

Results from our GMMs suggest that bulges host the oldest stellar populations and exhibit the lowest $z=0$ SFRs. They have higher ex-situ fractions than discs, but lower ex-situ fractions than the IHL at fixed mass. The ex-situ fraction of bulges correlates strongly with halo mass,
increasing from about 10 per cent at $M_{200}\approx 10^{12}\,{\rm M_\odot}$ to about $\approx 70$ per cent at $M_{200}\approx 6\times 10^{13}\,{\rm M_\odot}$; for galaxies hosted by haloes of mass $M_{200}\approx 10^{13}\,{\rm M_\odot}$, roughly half of all their bulge stars were formed ex-situ.

Although the IHL is dominated by ex-situ stars at most masses, our analysis suggests that it also contains a non-negligible fraction of stars that were born in-situ (roughly 30 per cent at the galactic scale, and
$\approx 10$ per cent at the cluster scale). This goes against a common assumption that the IHL is comprised entirely of accreted stellar material \citep[e.g.][]{cooper_galactic_2010}. At $M_{200} < 10^{12}\, \rm{M}_\odot$, the IHL is systematically older, less star-forming, and comprised of more ex-situ stars than bulges. Above this mass scale, 
the median SFRs and ages of bulges and the IHL are practically indistinguishable. This is in qualitative agreement with observational results indicating that the stellar populations of the ICL and BCGs overlap significantly \citep{jimenez-teja_unveiling_2018}.

The median age of discs and bulges increases with increasing $M_{200}$, consistent with the concept of galaxy ``downsizing'' \citep[e.g.][]{neistein_natural_2006}. The half-mass ages of the IHL component do not vary significantly with halo mass, and have a median age of $\approx 10$ Gyr. There are hints of an upturn in the IHL ages for galaxies with $M_{200} < 10^{12}\,\rm{M}_\odot$, though this is roughly the mass scale below which the IHL of most galaxies are resolved by fewer than 1000 particles. We explore the relationship between the age of the IHL component, the IHL mass fraction, and galaxy morphology further in Section \ref{subsubsec:fihl}.

Finally, note that the median SFRs and half-mass ages of the various galactic components are in good agreement between Ref-L0100N1504 and 50-HiResDM.  These results corroborate and extend the findings of \citet{ludlow_spurious_2023}, and suggest that spurious heating does not affect the SFRs of simulated galaxies, or even the SFRs of their dynamically-distinct components. This is because, at the mass resolution of our simulations, gaseous baryons are largely unaffected by spurious heating because their radiative cooling timescale is shorter than their collisional heating timescale \citep{steinmetz_two-body_1997}.

\subsection{The structure of discs, bulges and the IHL}
\subsubsection{The size-mass relation for discs, bulges, and the IHL}
\begin{figure*}
    \centering
    \includegraphics{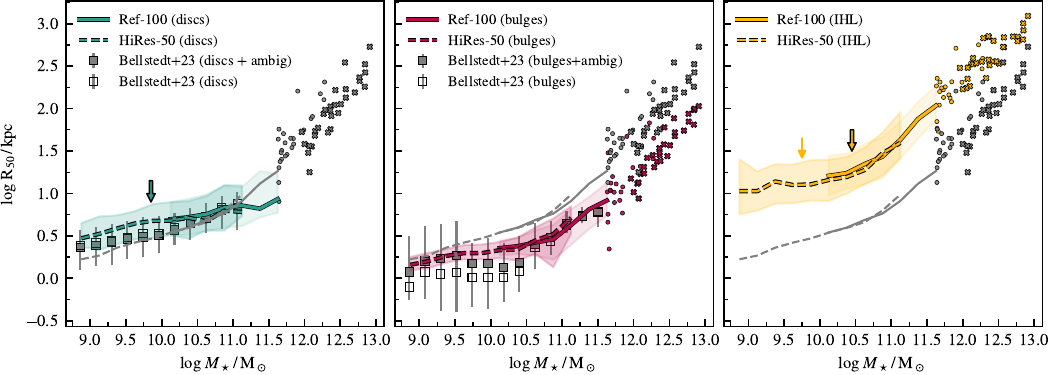}
    \caption{Projected half-mass size, $R_{50}$, plotted versus $M_{\star}$ for the disc, bulge, and IHL components of \eagle\ galaxies (left to right panels, respectively). The thick lines indicate the median relations, and the shaded regions enclose the interquartile scatter. Medians are plotted for mass bins that contain at least 10 galaxies (using dashed lines for 50-HiResDM and solid lines for Ref-L0100N1504); individual galaxies are plotted beyond this point. Arrows with (without) black boundaries indicate the halo mass above which each component is resolved by more than 1000 (100) stellar particles for at least 90 per cent of galaxies. The grey lines show the median $R_{50}-M_{\star}$ relations for the whole galaxy, and are repeated in each panel for comparison. Grey squares and error bars show a sample of observed half-mass sizes derived in \protect\cite{bellstedt_resolving_2023}.}
    \label{fig:r50}
\end{figure*}

Fig. \ref{fig:r50} plots the $R_{50}-M_\star$ relations obtained from our simulations. Note that $R_{50}$ is the projected half-stellar mass radius of the galaxy, which we obtain by multiplying its 3-dimensional half-mass radius by 0.75. For comparison, we also include the half-mass radii of discs and bulges obtained for GAMA galaxies as described in \cite{robotham_profuse_2022} (including additional galaxies from \citealt{bellstedt_resolving_2023}). Recall that our structural decomposition method makes use of the energies and angular momenta of stellar particles, quantities that are observationally inaccessible for stars in galaxies outside of the MW. \cite{robotham_profuse_2022}  and \cite{bellstedt_resolving_2023} estimate component sizes based on a simultaneous spectral and photometric decomposition of the observed systems. Despite these differences, the comparison serves as a useful test of our method. 

As above, results from
Ref-L0100N1504 are shown using solid lines, results from 50-HiResDM using dashed lines, and C-\eagle~ galaxies using crosses. Different panels
show results separately for discs (left), bulges (middle), and the IHL (right). For comparison, the grey lines in each panel show the size-mass relations for the total stellar component of each galaxy, regardless of morphological type. Open squares plot the median half mass radii for the decomposition presented in \cite{bellstedt_resolving_2023} for discs (left panel) and bulges (middle panel). Filled squares plot the same, but include ambiguous sources, where the component classification is uncertain \citep[see appendix D of][for details]{lagos_quenching_2023}.

The median disc sizes depend weakly on $M_{\star}$ across the mass range plotted, and agree well with the disc sizes obtained by \cite{bellstedt_resolving_2023}. For $M_{\star} < 10^{10}\, M_\odot$, our disc components are slightly larger than observed discs, though both exhibit similar scatter For $M_{\star} \lesssim 10^{10.5}\,{\rm M_\odot}$, bulge sizes also depend weakly on $M_{\star}$, but the relation steepens at higher masses where spheroids begin dominate our sample. The median bulge sizes are in excellent agreement with the observations across the mass range plotted, although the scatter in the simulated bulge sizes is lower. This may be due to the different stellar mass distributions between our sample and \cite{bellstedt_resolving_2023}: we impose a limit of $M_{200} > 10^{11.2}\,\rm{M}_\odot$ on our simulated sample, whereas the observed sample is complete down to $M_\star \approx 10^{9}\, \rm{M}_\odot$. Note that discs are larger than bulges by roughly 0.2 dex at a given mass, in agreement with other observational results \citep[e.g.][]{lange_galaxy_2016, robotham_profuse_2022}. Like bulges, the median size of the IHL component also exhibit an upturn at $M_{\star} \gtrsim 10^{10.5}\,{\rm M_\odot}$.

Finally, note the good agreement between the various size-mass relations obtained from Ref-L0100N1504 and 50-HiResDM.
Although we employed lower limits on $M_{200}$ for these runs such that the effects of spurious heating are negligible at (and above) 
the half-mass radii of {\em all} stellar particles, it is encouraging that good convergence is also obtained for the sizes of dynamically distinct
discs, bulges, and the IHL.

\subsubsection{The IHL transition radius}\label{subsubsec:regions}
\begin{figure}
    \centering
    \includegraphics{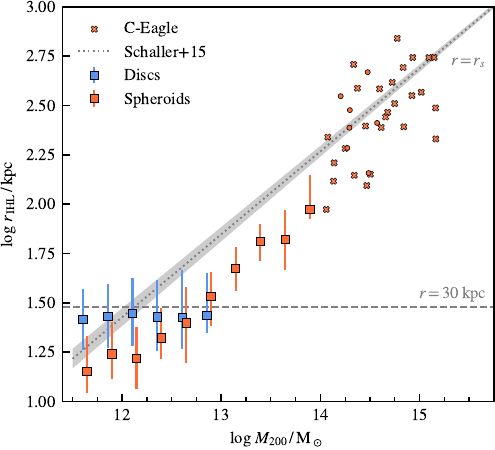}
    \caption{The IHL transition radius, \rtrans, plotted as a function of halo mass. Median values are shown separately for discs (blue squares) and spheroids (orange squares), and error bars represent the interquartile scatter. We show individual galaxies as circles for mass bins that contain fewer than 10 galaxies; crosses show galaxies in the C-\eagle~ sample. The median $r_s-\rm{M_{200}}$ relation for \eagle\, is plotted as a grey dotted line and was taken from \protect\cite{schaller_baryon_2015}.}
    \label{fig:ihl_trans}
\end{figure}

At what radius does the IHL begin to dominate the stellar mass of a galaxy? We denote this radius \rtrans\ and explore below how it depends on
halo mass and galaxy morphology.

To compute \rtrans, we first construct spherically-averaged density profiles for the different structural components of each galaxy,
using 36 equally-spaced bins in $\log(r)$ that span the range $-0.5 < \log(r/\rm{kpc}) < 3$. We then interpolate the profiles to determine
the radius at which the density of the IHL exceeds the density of the remaining stellar material. This procedure yields $r_{\rm IHL}$
values for $\approx 97$ per cent of our galaxy sample. Galaxies for which $r_{\rm IHL}$ could not be determined are usually low-mass, disc galaxies
with low IHL fractions. For such systems, sampling noise in the stellar halo makes it difficult to determine an accurate value of $r_{\rm IHL}$. For
that reason, in the remainder of this section, we only consider galaxies with $M_{200} \geq 10^{11.5}\, \rm{M}_\odot$, for which $r_{\rm IHL}$ was
reliably determined.

In Fig. \ref{fig:ihl_trans}, we plot \rtrans\ versus $M_{200}$ for galaxies identified in Ref-L0100N1504. Blue points show results for our sample of disc galaxies, and orange points show spheroids. The shape of the \rtrans$-\,M_{200}$ relation differs from the size-mass relations of the individual structural components of galaxies plotted in Fig.~\ref{fig:r50}. Specifically, below $M_{200} \approx 10^{12.8}\,$\msol, \rtrans\ is largely independent of $M_{200}$ for discs, for which the IHL begins to dominate the stellar mass distribution at \rtrans $\approx 30\, \rm{kpc}$ (shown as a horizontal dashed line in Fig.~\ref{fig:ihl_trans}). At these masses, the $r_{\rm IHL}$ values of discs are larger than for spheroids, and for the latter \rtrans\ increases proportional to $M_{200}$.

Although $r_{\rm IHL}$ marks the radius where the IHL begins to dominate the stellar mass of a galaxy, the typical fraction of the IHL within that radius can be quite large. For example, roughly 63 (56) per cent of the IHL mass of discs (spheroids) hosted by haloes with $M_{200}\approx 10^{12}\,{\rm M_\odot}$ lies within $r_{\rm IHL}$. And beyond $r_{\rm IHL}$,
only about 77 percent of the stellar mass is associated with the IHL (for discs; for spheroids it is 98 per cent). Although the exact values differ as a function of halo mass, our results suggest that there is significant overlap  in the spatial distribution of the different galaxy components, and that \rtrans\ (or any other characteristic radius, or fixed spherical aperture) is unlikely to accurately distinguish the IHL from the other components of a galaxy. 
Defining the IHL as such likely excludes a significant fraction of the total IHL mass (and includes non-negligible contributions to the IHL from bulge or disc stars) and potentially biases estimates of IHL properties.

The strong mass dependence of \rtrans\ at $M_{200}\gtrsim 10^{13}\,{\rm M_\odot}$ differs from the findings of \cite{contini_transition_2022},
who found that $r_{\rm IHL} \approx 60\,{\rm kpc}$ at the group and cluster scale. This discrepancy is perhaps due to the different theoretical approach to the problem (they used a semi-analytic model that implicitly assumes that the IHL follows a NFW profile, albeit one that is more concentrated than the surrounding DM halo by a factor of 3), but may also be due to  our different definitions of $r_{\rm IHL}$. Specifically, \cite{contini_transition_2022} identify $r_{\rm IHL}$ with the radius at which the IHL contributes 90 per cent of the total stellar mass, whereas in our definition it contributes 50 per cent. Regardless of these differences, the different mass dependence of \rtrans\ for disc galaxies above and below $M_{200} \approx 10^{13} {\rm M_\odot}$ seen in Fig.~\ref{fig:ihl_trans} indicates that the IHL in groups and clusters likely differs from the IHL of discs at the galaxy scale. A comprehensive analysis of the structure of the IHL, how it varies with mass, and its relationship to galaxy assembly histories warrants further investigation, which we defer to future work.
    
\cite{chen_sphere_2022} showed that the IHL transition radii inferred from photometric decomposition of stacked images of galaxy clusters in the Sloan Digital Sky Survey \citep{rykoff_redmapper_2014} are comparable to the characteristic scale radii (inferred from weak lensing) of their surrounding DM haloes. The grey line plotted in Fig. \ref{fig:ihl_trans} shows the best-fitting $r_s - M_{200}$ relation proposed by \cite{schaller_baryon_2015},
which was obtained by fitting NFW profiles to the average density profiles of relaxed haloes in \eagle. At the group and cluster scale, \rtrans\ exhibits a similar dependence on $M_{200}$ to that of $r_s$. Furthermore, above $M_{200}\approx 10^{14}\,$\msol, we find  \rtrans$\approx r_s$, although with considerable scatter. We did not, however, find a correlation between \rtrans\ and $r_s$ among individual systems, i.e. the scatter in \rtrans\ at fixed $M_{200}$ cannot be attributed to differences in halo concentrations. The apparent similarity between \rtrans\ and $r_s$ may be coincidental.

Finally, note that the \rtrans\ values we obtain at cluster mass scales are consistent with observational results  suggesting that the sphere of influence of BCGs can extend to $\approx200\, \rm{kpc}$ \citep{chen_sphere_2022}, which is significantly larger than the fixed spherical apertures often used in theoretical work \citep[e.g. 30 kpc;][]{montenegro-taborda_growth_2023}.

\subsection{Variation of the IHL fraction with host galaxy properties}\label{subsubsec:fihl}
\begin{figure*}
    \centering
    \includegraphics{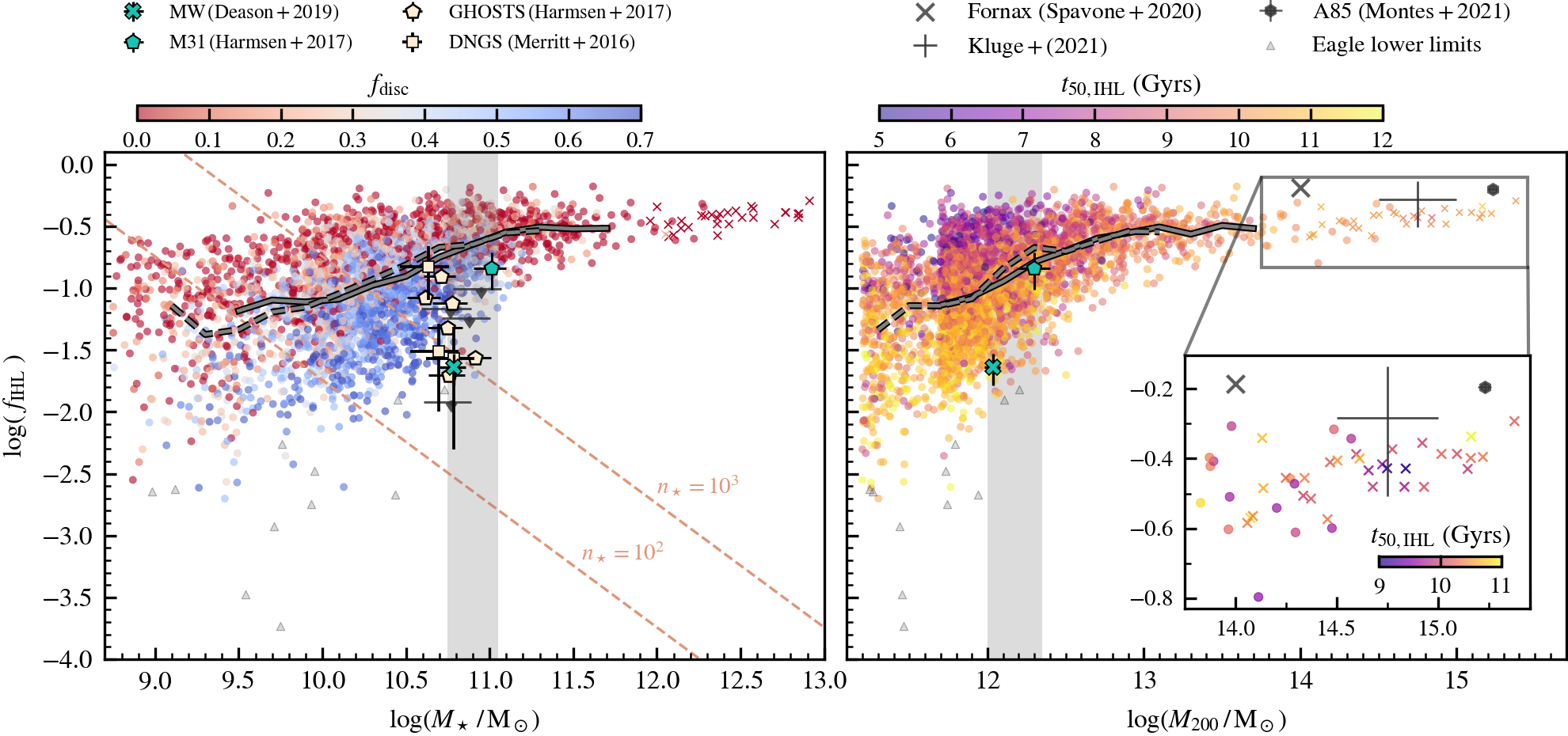}
    \caption{Left panel: \fihl\ as a function of stellar mass. Galaxies in Ref-L0100N1504 and 50-HiResDM are plotted as circles and are colour coded by \fdisk~ (those without an IHL component are plotted as grey triangles using the fraction of stellar mass at $r > 100\,{\rm kpc}$ as lower limit for \fihl). The median relations for the 50-HiResDM and Ref-L0100N1504 simulations are plotted in dashed and solid grey lines, respectively. C-\eagle\ galaxies are plotted using crosses. Observational results from the GHOSTS survey are plotted as beige pentagons with black boundaries \protect\citep{harmsen_diverse_2017}; results from the Dragonfly Nearby Galaxies Survey (DNGS; \citealt{merritt_dragonfly_2016}) are plotted as beige squares (the IHL fraction for these galaxies was estimated using the approach of \citealt{harmsen_diverse_2017}). Results for the MW \citep{deason_total_2019} and M31 \citep{harmsen_diverse_2017} are plotted using a green cross or pentagon, respectively. The observed IHL fractions of disc-dominated galaxies, where required, have been expressed relative to the total stellar mass in the central galaxy (see text for details). Right panel: \fihl\ as a function of $M_{200}$. \eagle\ galaxies are colour coded by the median stellar age of the IHL ($t_{50, {\rm IHL}}$). Observational results are shown for the Fornax cluster \protect\citep{spavone_fornax_2020}, A85 \protect\citep{montes_buildup_2021}, and a compilation of 170 local clusters by \protect\cite{kluge_photometric_2021}. Halo mass estimates for the MW and M31 were taken from \protect\cite{shen_mass_2022} and \protect\cite{fardal_inferring_2013}, respectively. The inset panel zooms in on \fihl\ for haloes with $\rm{M_{200}} > 10^{13.8}\,\rm{M_\odot}$, for which the colour bar has been appropriately rescaled.}
    \label{fig:fihl}
\end{figure*}

In the left panel of Fig. \ref{fig:fihl}, we plot \fihl\ versus $M_\star$ and compare our simulation results  to values obtained from observations of nearby disc galaxies (observational data were taken from Table 4 of \citealt{harmsen_diverse_2017}; our simulated galaxies are colour-coded by their disc mass fractions). The MW and M31 are shown using green symbols, and the rest are plotted in beige (with black edges). For the observed discs, we have included the contribution of the IHL to the total stellar mass.

In the right panel of Fig. \ref{fig:fihl}, we plot \fihl\ versus $M_{200}$ and compare with observed IHL fractions for a few galaxy clusters,\footnote{Observed IHL fractions at the group and cluster scale are typically expressed as light fractions. Here, we assume a constant mass-to-light ratio when comparing to our simulation results.} as well as for the MW and M31 (the virial masses for the MW and M31 were taken from \citealt{shen_mass_2022} and \citealt{fardal_inferring_2013}, respectively;
in this case, our simulated galaxies have been coloured according to the half mass ages of their stellar IHL particles). Note that at the galaxy cluster scale, it is common for IHL fractions to be expressed relative to the total stellar mass bound to the halo, inclusive of satellite galaxies. For our analysis, we only include observational data for which the contribution from satellite galaxies was excluded from the reported IHL fractions.

In both panels, solid and dashed grey lines show the running medians obtained from Ref-L0100N1504 and 50-HiResDM, respectively. The median \fihl\ values are in good agreement between the two simulations at all mass scales plotted, suggesting that our IHL mass estimates are robust to the spurious heating of stellar particles by DM particles. For simulated galaxies with no discernible IHL, we instead plot $f_{>100\,\rm{kpc}}$ (grey triangles in both
panels of Fig. \ref{fig:fihl}), which can be interpreted as a lower limit on their IHL fractions (see Fig. \ref{fig:fihl_comp}). 

The left panel of Fig. \ref{fig:fihl} shows that there is some overlap in the
\fihl\ values obtained for our simulated galaxies and from observations (although the simulated galaxies are biased toward slightly higher \fihl, on average). The IHL fractions of simulated galaxies measured using aperture-based methods \citep[e.g.][]{pillepich_first_2018, elias_stellar_2018} or 6D phase-space information \citep[e.g.][]{canas_stellar_2020} are also typically higher than observed IHL fractions. This suggests that the IHL of simulated galaxies may genuinely outweigh that of observed ones, although
it has been noted that \fihl\ values derived from single-band photometric data (such as those in \citealt{merritt_dragonfly_2016}) are likely lower limits on the true IHL fractions \citep{sanderson_reconciling_2018}.
We also stress that the observed IHL fractions plotted in the left panel of Fig. \ref{fig:fihl} correspond to disc-dominated galaxies, and more closely coincide with the IHL fractions of disc-dominated galaxies in our simulations (i.e. those corresponding to the blue coloured points).

The left panel of Fig. \ref{fig:fihl} also shows that there is a large diversity in the IHL fractions of simulated galaxies at fixed stellar mass. For those that span the stellar mass range of the MW and M31 (i.e. the vertical shaded region in the left-hand panel of Fig. \ref{fig:fihl}), the rms scatter (in $\log\,f_{\rm IHL}$) about the median IHL fraction is about $0.5$ dex, corresponding to a factor of $\approx 3$ in IHL mass. The scatter, however, clearly correlates with galaxy morphology: at fixed $M_\star$, galaxies with higher IHL fractions tend to have lower disc fractions, and vice versa.

The colour coding of points in the right-hand panel of Fig. \ref{fig:fihl} shows that there is also a correlation between the IHL mass fraction and its half mass age, $t_{\rm{50, IHL}}$, at least among low-mass galaxies (i.e.
$M_{200}\lesssim 10^{12.5}\,{\rm M_\odot}$). Note, however, that the correlation between $t_{\rm{50, IHL}}$ and \fihl\ disappears at high masses (see inset in the right-hand panel Fig.~\ref{fig:fihl}). At the scale of galaxy clusters, the IHL fractions obtained from our simulations approach a
constant value of \fihl$\approx0.37$, systematically lower than the values reported by \cite{kluge_photometric_2021}, who found \fihl$=0.52$ using a double Sersic decomposition. 

The fact that the IHL of disc dominated galaxies tends to be older and less massive than that of spheroids of similar stellar mass can be interpreted as follows. Galaxies with high disc mass fractions are unlikely to have
experienced recent disruptive mergers that could potentially contribute to the growth of their IHL. As a result, what IHL they do possess tends to be older than that of spheroidal galaxies, which are more likely to have experienced recent mergers. This interpretation is supported by the conclusions of \cite{deason_total_2019}, who showed that the low IHL mass of the MW can be largely explained by an ancient ($\approx 10\, \rm{Gyr}$) merger event with a massive progenitor whose stellar mass now dominates its IHL.

The diagonal dashed lines in the left-hand panel of Fig. \ref{fig:fihl} highlights IHL fractions corresponding to roughly 100 and 1000 stellar particles. Clearly, the stellar haloes of many of the low-mass galaxies in our simulations (i.e. those with $M_\star\lesssim 10^{10}\,{\rm M_\odot}$) are only resolved by a few 10s to a few 100s of stellar particles. Assessing the properties of their IHL components, such as their shape or spatial and kinematic structure, will likely require simulations that reach higher baryonic mass resolution than our runs.
\section{Conclusions and outlook}\label{sec:conc}

We used Gaussian Mixture Models (GMMs) to decompose the structural components of central galaxies identified in the $z=0$ output of the \eagle\ simulation. Most of our analysis was based on galaxies identified in Ref-L0100N1504, i.e. the 100 cubic Mpc flagship simulation of the \eagle\ Project \citep{schaye_eagle_2015}, which we supplemented with 30 galaxy clusters from the C-\eagle\ Project \citep{bahe_hydrangea_2017, barnes_cluster-eagle_2017}, and several hundred from 50-HiResDM \citep{ludlow_spurious_2023}. The latter run used the same numerical and subgrid model parameters as Ref-L0100N1504, but was was carried out in a smaller volume (50 cubic Mpc) and with seven times higher mass resolution in the DM component. Combined, these simulations allowed us to study the discs, bulges, and intra-halo light (IHL) of galaxies across a range of environments and over four orders of magnitude in halo mass ($10^{11.2}\,{\rm M_\odot}\leq M_{200} \leq 10^{15.4}\, {\rm M_\odot}$). Our main results are summarised below.

\begin{itemize}

\item Our galaxy decomposition technique is robust to small variations in $n_c$, i.e. the number of Gaussian distributions used by the GMM to isolate kinematically distinct galaxy components (see Section \ref{sec:decomp}). The stellar mass fractions assigned to the disc, bulge, and IHL components of galaxies are, on average, independent of $n_c$ provided $8\lesssim n_c \leq 15$ (although there is some variation on the level of individual galaxies). For most of our analysis we used $n_c=12$, which resulted in disc fractions that correlate strongly with alternative kinematic morphology estimators, such as $\kappa_{\rm co}$ (\citealt{correa_relation_2017}; see Fig. \ref{fig:fdisc_comp}). We also reproduce the morphology-dependence of the stellar-to-halo mass relation reported in previous observational and theoretical work (Fig. \ref{fig:fstar_morph}). 

\item Of the 3342 galaxies in our sample, roughly 71 per cent were classified as discs (i.e. have a non-zero disc mass fraction based on the $n_c=3$ GMM; see Section \ref{subsec:decomp_expl}), the remainder were classified as spheroids. We find that 34 per cent of the disc population are disc dominated (i.e. have $f_{\rm disc}>0.5$). Disc galaxies primarily occupy haloes with virial mass $M_{200}\lesssim 10^{12.5}\,{\rm M_\odot}$, whereas spheroids dominate in higher-mass haloes. Less than 0.5 per cent of galaxies in our sample possess no IHL component; these are primarily low-mass, disc-dominated systems (i.e. they typically have disc mass fractions $f_{\rm disc}>0.5$).
  
\item The basic properties of the disc and bulge components of galaxies are consistent with well-established observed trends: discs host younger stellar populations than bulges, have higher star formation rates, and, on comparable mass scales, are systematically larger than bulges (as quantified by their stellar half-mass radii). The half-mass radii of discs and bulges are in good agreement with observational results presented in \cite{robotham_profuse_2022} and \cite{bellstedt_resolving_2023} (see Fig. \ref{fig:r50}). Discs also contain a smaller contribution from ex-situ stars than bulges, by roughly a factor of 2 (see Fig. \ref{fig:meds}).
  
\end{itemize}

The primary aim of our work, however, was to study the properties of the IHL components of $z=0$ galaxies using a consistent methodology for IHL identification, and across a broad range of galaxy and halo masses. The results listed above give us confidence that our galaxy decomposition technique is sensible and robust, and that our IHL definition is on firm footing. The main insights into the IHL components of galaxies that our work provides are summarised as follows:

\begin{itemize}

\item Compared to discs and bulges, the IHL component of galaxies contains the highest fraction of ex-situ stars, which contribute roughly 70 per cent of the IHL mass at $M_{200}\approx 10^{12}\,{\rm M_\odot}$ and about 90 per cent at $M_{200}\approx 10^{14}\,{\rm M_\odot}$. The average SFR of the IHL (averaged over a lookback time of 500 Myr) is lower than that of bulges and discs. The IHL is, on average, composed of older stellar populations than discs and bulges at $M_{200} < 10^{12.5}\,{\rm M_\odot}$; above this mass scale the ages of bulges and the IHL overlap (Fig.~\ref{fig:meds}). At all mass scales, the IHL component is more extended than the disc or bulge component (as quantified by the half stellar mass radius of each component; Fig.~\ref{fig:r50}). The SFRs, ages, and sizes of the disc, bulge, and IHL components studied in this work are converged between the Ref-L0100N1504 and 50-HiResDM simulations, indicating that our results are robust to the effects of spurious collisional heating \citep[see][for details]{ludlow_spurious_2023}.
  
\item The fraction of mass assigned to the IHL, \fihl, increases with increasing stellar and halo mass (although slightly), but at fixed mass exhibits large scatter. For MW mass galaxies (i.e. $M_\star\approx 10^{10}\,{\rm M_\odot}$ or $M_{200}\approx 10^{12}\,{\rm M_\odot}$) we find \fihl $\approx 0.08$, which increases to \fihl $\approx 0.37$ at the group and cluster scale (i.e. $M_{200}\gtrsim 10^{13}\,{\rm M_\odot}$). The IHL fractions we obtain from our GMMs are in broad agreement with observed values obtained for disc and BCGs (Fig. \ref{fig:fihl}; although several nearby disc galaxies have systematically lower IHL fractions than discs in our simulations).

\item At halo masses $M_{200} \lesssim 10^{12.5} \rm{M_\odot}$, the scatter in \fihl\ is closely connected to the kinematic morphology of a galaxy: at a fixed halo mass, disc dominated galaxies have lower IHL fractions than spheroidal galaxies. The IHL fraction also correlates (albeit weakly) with the median age of the IHL (Fig. \ref{fig:fihl}) in such a way that the IHL of disc-dominated galaxies is systematically older than the IHL of spheroids of the same mass. This supports the idea that most disc-dominated galaxies have not undergone any recent mergers that could have increased their IHL fractions, and that, more broadly, the IHL fraction of a galaxy holds valuable information about its assembly history.

\item For halo masses $M_{200} \gtrsim 10^{13}\,{\rm M_\odot}$, the various correlations between the IHL mass fraction, its age, and galaxy morphology no longer hold. This may be due to the lack of diversity in galaxy morphologies at these mass scales (Fig. \ref{fig:fstar_morph}) or due to similarities in the merger histories of massive haloes. At these mass scales, galaxies are typically spheroidal and both the bulge and IHL components are dominated by ex-situ stars. It is not clear whether a meaningful distinction exists between the bulge and IHL components of galaxies in massive haloes, since both components exhibit similar stellar populations (Fig. \ref{fig:meds}).

\item Finally, we explored how the IHL transition radius, \rtrans (defined as the radius beyond which the stellar density profile of the IHL dominates over that of the inner galaxy), depends on halo mass and galaxy morphology. For disc galaxies, we found that \rtrans$\approx 30\,{\rm kpc}$. For spheroids, \rtrans\ depends strongly on $M_{200}$, increasing from  $r_{\rm IHL}\approx 20 \,{\rm kpc}$ at $M_{200}\approx 10^{12}\,{\rm M_\odot}$ to $r_{\rm IHL}\approx 300 \,{\rm kpc}$ at $M_{200}\approx 10^{15}\,{\rm M_\odot}$ (Fig. \ref{fig:ihl_trans}). Our methodology also predicts significant spatial overlap between galaxy components, implying that \rtrans\ (or any other spherical aperture) cannot be used to distinguish the IHL from the other structural components of a galaxy. 

\end{itemize}

We believe our results provide a sensible assessment of some of the basic properties of the disc, bulge, and IHL components of simulated galaxies. We plan to leverage the galaxy decomposition technique introduced in this work to explore properties of the progenitors of the IHL (as well as the progenitors of the ex-situ components of discs and bulges) and how they depend on halo mass.

Although our simulated galaxy sample spans a large range of halo masses ($10^{11.2}\,{\rm M_\odot}\leq M_{200} \leq 10^{15.4}\, {\rm M_\odot}$), massive galaxy clusters are sparsely sampled, and the IHL of low-mass galaxies ($M_{200}\lesssim 10^{12}\,{\rm M_\odot}$) is poorly resolved (typically containing only a few hundred to a few thousand stellar particles). Future work on the subject may benefit from recent large volume cosmological simulations that provide much larger samples of massive clusters \citep[e.g.][]{pakmor_millenniumtng_2023, kugel_flamingo_2023, schaye_flamingo_2023-2}. Likewise, cosmological simulations with higher baryonic resolution than \eagle\ will be useful for exploring the structure of the IHL of low-mass galaxies, which is crucial if we wish to properly interpret the low IHL fraction of the MW and other nearby galaxies, and to properly place their formation histories in a wider cosmological context \citep[e.g.][]{evans_how_2020}. Such simulations will also enable investigations into the IHL of dwarf galaxies, which, due to their low surface brightness, are difficult to study observationally but are potentially powerful probes of dark matter models \citep{deason_dwarf_2022}.

\section*{Data Availability}
The {\sc EAGLE} simulations are publicly available; see \citet{mcalpine_eagle_2016,the_eagle_team_eagle_2017} for how to access {\sc EAGLE} data.
Any additional data used in this work can be made available upon reasonable request. The decomposition code is publicly available.\footnote{\href{https://github.com/katyproctor/decomp}{https://github.com/katyproctor/decomp}}

\section*{Acknowledgements}
We thank the anonymous referee for a thoughtful report that improved the quality of the paper. We also wish to thank Joel Pfeffer for providing ex-situ particle classifications, and Matthieu Schaller for providing best-fitting NFW concentrations for the Ref-L0100N1504 simulations. KLP thanks Annette Ferguson, Azi Fattahi, and Alexander Knebe for useful discussions and thanks the University of Edinburgh for supporting a productive research visit. KLP acknowledges support from the Australian Government Research Training Program Scholarship. CL has received funding from the Australian Research Council Centre of Excellence for All Sky Astrophysics in 3 Dimensions (ASTRO 3D), through project number CE170100013, and the Australian Research Council Discovery Project (DP210101945). ADL acknowledges
financial support from the Australian Research Council through their Future Fellowship scheme (project number FT160100250). ASGR acknowledges funding by the Australian Research Council (ARC) Future Fellowship scheme (FT200100374, `Hot Fuzz').
This work made use of the supercomputer OzSTAR which is managed through the Centre for Astrophysics and Supercomputing at Swinburne University of Technology. This supercomputing facility is supported by Astronomy Australia Limited and the Australian Commonwealth Government through the national Collaborative Research Infrastructure Strategy (NCRIS). 
We acknowledge the Virgo Consortium for making
their simulation data available. The \eagle\ simulations were performed using the DiRAC-2 facility at
Durham, managed by the ICC, and the PRACE facility Curie based in France at TGCC, CEA, Bruyeres-le-Chatel.
The C-\eagle\ simulations were in part performed on the German federal maximum performance computer ``HazelHen'' at the maximum performance computing centre Stuttgart (HLRS), under project GCS-HYDA / ID 44067 financed through the large-scale project ``Hydrangea'' of the Gauss Center for Supercomputing. This work has benefitted from the following public {\sc python} packages: {\sc pandas} \citep{mckinney_data_2010}, {\sc scipy} \citep{virtanen_scipy_2020}, {\sc numpy} \citep{harris_array_2020}, and {\sc matplotlib} \citep{hunter_matplotlib_2007}.

\bibliographystyle{mnras}
\bibliography{paper} 


\appendix

\bsp	
\label{lastpage}
\end{document}